# A Thin Flexible Acoustic Transducer with piezoelectric-actuated microdomes for Underwater Communication

Rong Fu[1,3], Xinyu Zhang[2,3], Cheng-Hao Yu[1], Kai Liu[1], Tauhidul Haque[1], Leixin Ouyang[1], Mark Ming-Cheng Cheng[1]

1. Department of Electrical and Computer Engineering, The University of Alabama, Tuscaloosa, Alabama, 35487, US

2. The Department of Computer Science, The University of Alabama, Tuscaloosa, Alabama, 35487, US

3. Denoting the co-first authors with the same contributions in this paper.

Abstract

This paper presents a flexible thin-film underwater transducer based on a mesoporous PVDF membrane embedded with piezoelectrical-actuated microdomes. To enhance piezoelectric performance, ZnO nanoparticles were used as a sacrificial template to fabricate a sponge-like PVDF structure with increased β-phase content and improved mechanical compliance. The device was modeled using finite element analysis and optimized through parametric studies of dome geometry, film thickness, and dome size. Acoustic performance was evaluated through underwater testing, demonstrating high SPL output and reliable data transmission even at low drive voltages. The proposed transducer offers a lightweight, low-cost, and energy-efficient solution for short-range underwater communication in next-generation Ocean IoT systems.

## I.    Introduction

The Internet of Things (IoT) technology [1] has rapidly devolved in recent years, finding applications across various domains, including smart transportation [2], intelligent healthcare [3], industrial automation [4], forest monitoring [5], etc. While IoT has been extensively studied in terrestrial environments, its deployment in the ocean remains a significant challenge due to the complexity and harsh conditions of the marine environment [6]. Ocean IoT (OIoT) is an advanced network that leverages wireless communication to detect and identify underwater objects while monitoring the underwater environments [7]. As a key technology in the OIoT, underwater acoustic (UWA) communication is the most reliable method for long-distance transmission underwater [8], and it has been widely applied in various underwater-related fields. Compared to the other two mainstream technologies (Radio Frequency (RF) and Optical Waves), UWA propagates more effectively and reliably in underwater environments. However, it is constrained by low bandwidth, high latency, and is highly susceptible to multipath effects and Doppler shifts [9]. Today, different varieties of transducers are available and widely used in industrial and commercial applications. Nevertheless, many of these commonly used transducers are not well-suited for compact application platforms such as lightweight unmanned underwater vehicles (UUVs) or autonomous underwater vehicles (AUVs) [10] due to their large size, high cost, and substantial energy consumption. While ceramic and single-crystal piezoelectric materials generally exhibit high piezoelectric coefficients [11], [12], their rigid and brittle nature, along with the requirement for high-temperature processing, makes them unsuitable for applications in flexible, low-cost ultrathin-film transducers. Thereby, flexible thin-film transducers based on polymers with intrinsic or artificial piezoelectricity are attracting great attention because of their light weight, flexibility, scalability, non-intrusive installation, and the potential for low-cost, high-throughput fabrication [13]. Among piezoelectric polymers, polyvinylidene fluoride (PVDF) stands out as a strong candidate for underwater transducer design due to its low acoustic impedance, mechanical flexibility, wide bandwidth, and low fabrication cost [14]. Furthermore, it facilitates impedance matching with water, enabling efficient energy transfer between the transducer and the water [15].

In this paper, we present a flexible thin-film transducer based on an enhanced PVDF membrane that operates through the vibration of active array microstructures in a piezoelectric layer, which is sandwiched between two thin perforated silicone layers, which is shown in Fig.1. Spherical diaphragms serve as the microstructures in our design, precisely shaped and dimensioned through microfabrication. The electrical excitation induces vibration in the densely distributed microstructures, which can lead to outstanding acoustic performance. We then integrate simulations to investigate the characteristics of PVDF under various geometric parameters and validate the results through a series of experiments. Furthermore, we set up a communication system to evaluate its performance in underwater environments.

The main contributions of this work can be summarized as follows.

1) In our design, we employ a novel sponge-like mesoporous piezoelectric PVDF thin film for underwater transducer developments. The mesoporous structure enhances the piezoelectric performance and thin film flexibility by inducing more formation of the β-phase (the active piezoelectric phase) through increased surface area and enhanced interactions during the fabrication.

2) The mesoporous PVDF thin film is fabricated by casting a mixture of PVDF solution and zinc oxide (ZnO) nanoparticles, which introduce porosity to modulate the mechanical properties and promote the formation of the piezoelectric β-phase.

3) In our design, the active PVDF thin film possesses a microscale thickness and incorporates an array of microdomes. These engineered microstructures enhance signal output and contribute to the transducer's mechanical stability and long-term durability.

The rest of this article is organized in the following. In Section II, we summarize and categorize the relevant works. In Section III, we present the theoretical modeling of our design. In Section IV, we provide detailed explanations and fabrication procedures of the proposed device. Then in Section V, we use the COMSOL platform to study the influence of geometric parameters on the acoustic performance of the flexible transducer. Subsequently, the related experiments and analysis are presented in Section VI. Finally, the conclusion of this article is organized in Section VII.

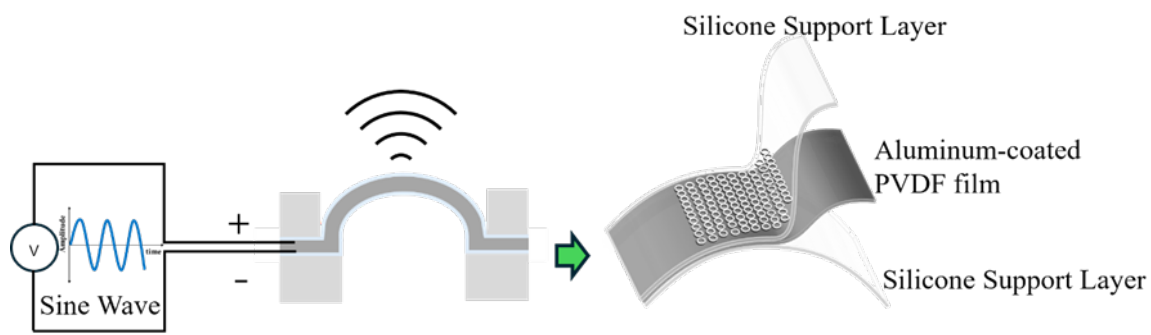

Fig.1 The structure of PVDF-based Transducer with microdomes

## II. Related Works

In this section, we present a brief review of previous works on the transducer design for underwater communication, mainly focusing on material selection and structural configurations. Moreover, we list some main differences between our work and other peer methods.

Underwater communication primarily relies on three key technologies: acoustic, optical, and radio frequency (RF) methods, each offering distinct advantages and limitations depending on the application and environmental conditions. The acoustic method has been widely investigated for underwater communication due to the favorable propagation characteristics of acoustic waves at low frequencies (up to tens of kHz). Unlike radio frequency (RF) and optical signals, which suffer from high attenuation and sensitivity to environmental conditions, acoustic waves offer significantly lower absorption, enabling long-distance signal transmission [16]. However, underwater acoustic communication still faces substantial challenges, including signal attenuation, ambient noise, Doppler shifts, multipath interference, and scattering effects from bubbles and suspended particles [17], all of which can degrade signal quality, reduce transmission range, lower data rates, and increase bit error rates—ultimately compromising the reliability and efficiency of underwater wireless communication systems.

To address these problems, numerous transducer designs have been proposed focusing on broadband performance, high operating frequency, and wide beam coverage. Reza Ghaffarivardavagh et al. [18] they presented an Ultra-wideband Underwater Backscatter ($U^2B$), which synthesizes different forms of resonance through the multi-layer (meta material) to overcome the bandwidth limitation while maintaining high efficiency. Yanjun Zhang et al. [19] demonstrated a high-frequency spherical-omnidirectional transducer composed of six spherically curved square 1-3 piezoelectric composite elements, achieving the transmitting voltage response (TVR) exceeding 158.3 dB at 1 m across the frequency range of 260-310 kHz with a peak value of 161.3 dB at 285 kHz. Jiuling Hu et al. [20] proposed a new type of flextensional transducer, composed of double mosaic piezoelectric ceramic rings and spherical cap metal shells, which achieves a maximum TVR of 147.2 dB and operates within a bandwidth of 1.5-4 kHz.

Recently, flexible transducers are increasingly favored for underwater communication applications due to their mechanical adaptability and integration potential with non-planar surfaces. PVDF stands out among piezoelectric polymers for its favorable acoustic impedance, piezoelectric response, and chemical stability in marine environments. One of the earlier studies, Minoru Toda et al. [21] introduced a wideband PVDF-based transducer using a coiled film coupled to a vibrating disc. This design achieves a broad response below 100 kHz, but lacks control over beam direction. M. Martins et al. [22] presented an optimization study for piezoelectric ultrasound emitter transducers for underwater communication, specifically investigating piston-type transducers in thickness mode. They compared single-layer and multilayer configurations of PVDF and PZT-5H materials, showing that the multilayer PVDF transducer offered better bandwidth and lower power consumption. More recently, Marcos S. Martins et al. [23] developed a wideband and wide beam PVDF acoustic transducer with a usable frequency range from 100 kHz to 1.5 MHz and a maximum TVR of 150 dB, making it suitable for short-range communication up to 15 meters.

For clarity, the comparison between our work and related works is listed in Table I. The existing research has made promising progress in transducer design. Compared with the existing designs, our design features a highly compact and ultra-thin PVDF configuration (3 cm × 3 cm × 0.0025 cm), offering sufficient response (108 dB) for short-distance communication (100 cm) at a low excitation voltage (20 V). While the output pressure is lower than that of traditional ceramic-based devices, the structure enables significant advantages in terms of flexibility, power efficiency, and miniaturization.

Table I. Comparisons between our work and existing work

| Design | Dimensions | Test Distance | Voltage | Maximum |
|--------|------------|---------------|---------|---------|

|  | (cm³) | (cm) | (V peak-to peak) | Response Results (dB) |
|---|---|---|---|---|
| [18] | 5.4(D)×4(H) | 400 | / | 29 |
| [19] | 5(D)×0.5(H) | 100 | 10 | 161.3 |
| [20] | 36(D)×10(H) | 500 | / | 147.2 |
| [22] | 1(D) ×0.011(H) | 3 | 10 | 147.96 |
| [23] | 1.7(L)×9.2(W)×0.011(H) | 100 | 20 | 152 |
| Our work | 3(L)×3(W)×0.0025(H) | 100 | 20 | 108 |

D: Diameter, H: Height, L: Length, W: Width.

## III. Theoretical Modeling

The vibration of the dome plays a crucial role in determining the acoustic performance of a transducer. When an AC voltage is applied across the electrodes of a piezoelectric dome, the resulting electric field generates strain within the material, causing the dome to oscillate in a rhythmic motion of expansion and contraction. This oscillation displaces the surrounding water, producing sound waves [24]. The structure of the dome and the relevant variables are illustrated in Fig. 2. To simplify the modeling, the horizontal components of the electric field are disregarded. It is assumed that the elastic and in-plane piezoelectric properties of the porous PVDF film are isotropic and unaffected by the thin metal electrodes deposited on the film. The film thickness is considered uniform across the dome, and voltage-induced change in thickness is neglected.

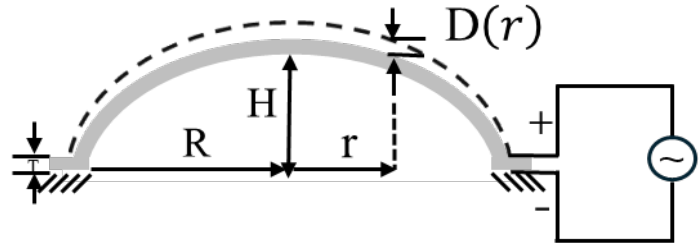

Fig.2 Schematic of a piezoelectric dome with corresponding geometric variables

Since the film thickness T is much smaller than the radius R (R/T>100), the shallow dome can be approximated as a spherical membrane with clamped edges when excited [24]. So the corresponding dome height can be calculated by

$$x(V) = \sqrt{\frac{2dV(R^2+H^2)}{T} + H_0^2} \quad (1)$$

Where d is the piezoelectric constant, V is the applied excitation voltage. R, H, and T represent the dome radius, dome central height, and the thickness of the piezoelectric membrane respectively. H0 denotes the dome's central height under the steady-state situation.

To describe the deflection of the piezoelectric dome, a simplified membrane model is employed [25]. The peak deflection value along the radial direction (with r=0 corresponding to the dome center), induced by the periodic voltage, can be obtained by

$$D(r) = \frac{x(V_m) - x(-V_m)}{2}\left[1 - \left(\frac{r}{R}\right)^2\right] \quad (2)$$

Where $V_m$ represents the amplitude of the excitation voltage, and r is the radial distance from the dome center to the deflection point.

For the ultrathin and flexible transducer, the thickness (at the microscale level) is significantly smaller than the acoustic wavelength within the frequency range of interest. When the flexible transducer is mounted on a rigid baffle and driven by a sinusoidal voltage, the acoustic pressure in the free field resulting from the dome's vibration can be calculated using the Rayleigh integral [26]. The Rayleigh integral is discretized since our transducer was fabricated based on microscale domes. Consequently, the acoustic pressure at the position $\boldsymbol{r}_d$ becomes a summation of the acoustic pressure generated by each dome at $\boldsymbol{r}_d$. The total acoustic pressure $P(\boldsymbol{r}_d)$ produced by the acoustic surface at a point located at $\boldsymbol{r}_d$ is given by

$$P(\boldsymbol{r}_d, f, V_n) = i2\pi\rho f^2 \sum_{n=1}^{N} \overline{w_n}(V_n) A_n e^{i\varphi_n} \frac{e^{i\frac{2\pi f}{c}|\boldsymbol{r}_d - \boldsymbol{r}'_{dn}|}}{|\boldsymbol{r}_d - \boldsymbol{r}'_{dn}|} \quad (3)$$

Where $\boldsymbol{r}_d$ is the vector from the designed origin to the observation point, $\boldsymbol{r}'_{dn}$ represents the vector from the origin to the location on the acoustic surface of the micro-dome $n$, $A_n$ and $\overline{w_n}$ are the location and average deflection(along the normal direction) of the micro-dome $n$, respectively. $c$ denotes the speed of sound in medium, $f$ is the excitation frequency, $V_n$ is the amplitude of the driving voltage applied to the dome $n$, and $\varphi_n$ characterizes the relative phase of vibration of the dome $n$.

The sound generation of the acoustic surface was characterized by driving all the microdomes in phase with the same voltage. The microscale domes on each sample we fabricated also have identical dimensions and thereby the same area and average deflection. The acoustic pressure can be converted into sound pressure level (SPL) by

$$SPL(\boldsymbol{r}_d, f, V_n) = 20Lg\frac{|P(\boldsymbol{r}_d, f, V_n)|}{P_r} \quad (4)$$

Where $P_r$=1μPa is the reference sound pressure underwater.

The resonance frequency of the domes is an important factor that affects the device's bandwidth and can be utilized to optimize performance at a specific frequency. To estimate the first resonance frequency of a shallow dome, it can be approximated as a flat circular thin film with clamped edges in a dynamic state. This dynamic model, developed in [27], provides the calculation for the first resonance frequency as follows

$$f_r = \frac{k_r}{2\pi}\sqrt{\frac{D}{ps}k_r^2 + \frac{T}{ps}} \quad (5)$$

Where $ps$ is the area mass density of the circular film, and $k_r$ is the wavenumber that is solved based on the clamped boundary conditions.

## IV.  Device Design and Fabrication

The structure of the ultrathin-film transducer is presented in Fig.1. The device consists of a piezoelectric layer with an array of micro-domes sandwiched between two layers of perforated silicone. To fabricate an ultrathin and flexible transducer, PVDF is chosen as the piezoelectric layer in our design since it has the advantages of low acoustic impedance, mechanical flexibility, larger bandwidth, and low-cost manufacturing. The top silicone layer is designed not only to protect the domes from abrasion and mechanical impact but also make enough space for micro-domes to vibrate. The bottom perforated silicone layer acts as a spacer to separate the domes, and forms spacious cavities for microstructures as well as reduces any influence from imperfect bonding and morphology of the surface on which the transducer is installed [24]. In this design, these convex microstructures can vibrate freely even if the transducer is bonded to a rigid object because they are not in contact with the mounting surface. The active microstructures of the PVDF film generate acoustic waves based on the d33 piezoelectric response. The PVDF film is poled along the thickness direction, which means that the electric field along the thickness direction of the PVDF film induces in-plane strain and causes the piezoelectric domes with fixed edges to expand and contract periodically by applying an AC voltage. The microdome vibration thereby moves the surrounding medium to generate sound. These locally curved domes on the piezoelectric films can enhance sound generation compared to the flat films [28,29], enabling the whole transducer to emit and transmit high-sensitivity sound in different application scenarios. Besides, the acoustic impedance of PVDF is closer to that of water compared to ceramic materials typically used in ultrasonic transducers, making PVDF transducers transmit and receive signals across a wide frequency range with reduced reflection and loss at the interface.

The fabrication process of the ultrathin and flexible transducer is shown in Fig.3. The process can be divided into two parts: PVDF film preparation and vacuum-induced embossing process.

During PVDF film preparation, PVDF powder was dissolved in *N, N*-dimethylformamide (DMF) and stirred at 45 °C for approximately 5 hours. To enhance film porosity, ZnO nanoparticles (NPs) were added to the solution, followed by an additional 2 hours of stirring to ensure uniform dispersion. The size of ZnO NPs used in our work ranges from 35nm to 45nm. Further to obtain a uniform PVDF/ZnO NPs suspension, the mixture was treated in an ultrasonic bath for 1 hour. After this, PVDF/ZnO NPs suspension was screen printed on a glass substrate and dried in a vacuum oven for 15 hours at 60°C to remove residual DMF. HCl acid solution was selected to remove ZnO NPs, as PVDF is a hydrophobic polymer with excellent chemical stability against corrosive solvents including acids [30,31]. In this design, the films were soaked in 37 wt% HCl solution for 6 hours to etch, then the films were rinsed in deionized (DI) water. To investigate the effect of varying ZnO NPs loadings on the structural and functional properties of PVDF, three films were fabricated with ZnO contents of 40 wt%, 50 wt%, and 60 wt%, respectively. Then a comprehensive set of characterizations like nanoindentation, Fourier-transform infrared spectroscopy (FTIR), and scanning electron microscopy (SEM) was conducted. The related characterizations of the films were discussed in the former work [15], and among these three films, 50%wt PVDF film has the best piezoelectric properties. After the film fabrication, an optical microscope was used to observe its thickness and Thorlabs SD-OCT was employed to confirm the dome height. 120 nm aluminum was deposited on both sides of the porous PVDF films as electrodes via e-beam evaporation.

During the vacuum-induced embossing process, a PVDF film was adhered to the perforated acrylic board. The surface area of the film facing the through-holes deforms into spherical domes due to the pressure difference across the film. The diameter of the through-holes determines the base diameter of the domes, while the applied vacuum pressure governs their height [15]. The perforated acrylic board was then placed on the top of a funnel, and the vacuum was connected. In order to maintain the dome shape, the PVDF film was heated at 100 °C on the hot plate for 15 minutes under vacuum. Subsequently, it was cooled to room temperature for 15 minutes while the vacuum was on. The entire procedure was repeated four times to ensure the consistent and reliable formation of the dome structures. After embossing, two laser-cut perforated silicone layers were aligned and bonded to the film. Meanwhile, copper tapes were placed on both sides for electrical connection. To prepare the transducer for underwater communication, a 1.5 μm parylene layer was deposited via PDS 2010 Labcoter to serve as an insulator. The example device is shown in Fig.3 (c), and the scan image of domes are presented in Fig.3 (d).

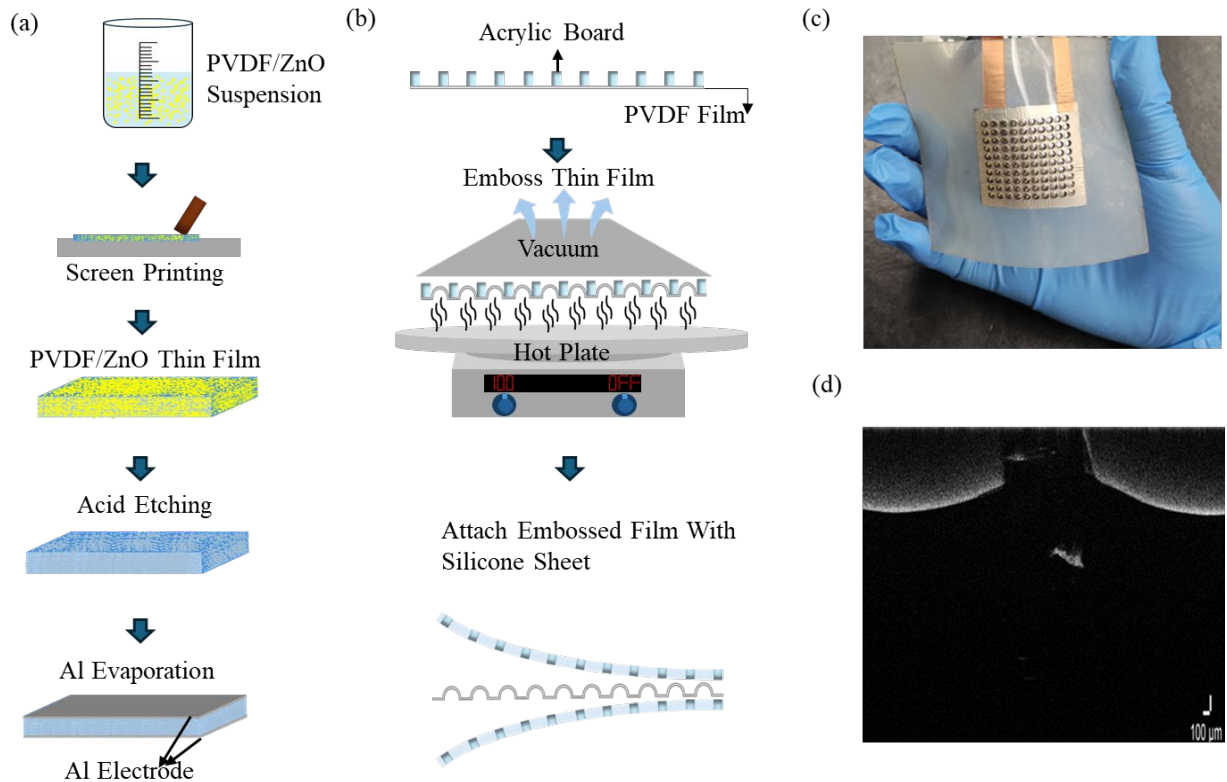

Fig.3 The fabrication procedure of ultrathin transducer. (a) PVDF film preparation. (b) Vacuum-induced embossing process. (c) Photograph of an ultrathin PVDF-based transducer. (d) OCT scan image of the fabricated domes.

## V. Finite Element Simulation

To explore the influence of geometric parameters on the acoustic performance of the flexible transducer, we utilize the COMSOL Multiphysics platform, which provides a robust environment for finite element analysis (FEA). First, 100 $\mu m$ PVDF films with two different structures (flat-shaped and dome-shaped) were simulated to evaluate and compare the mechanical performance of these two geometries. The corresponding results are presented in Fig. 4. For the flat-shaped structure, the simulation shown in Fig.4 (b) exhibits minimal displacement across the surface under applied electric fields, indicating limited deformation capability. This suggests that flat geometry is less effective in displacing the surrounding medium and generating significant sound. Additionally, the eigenfrequency of the flat structure was found to be lower, as shown in Fig.4 (a), reflecting its rigid mechanical behavior with limited dynamic flexibility. In contrast, the dome-shaped structure achieves substantially larger displacements under the same driven condition, peaking at the dome's apex and thereby enhanced the sound generation. It also showed a higher eigenfrequency, illustrating its ability to resonate more effectively at higher frequencies, which is beneficial for sound generation in practical applications.

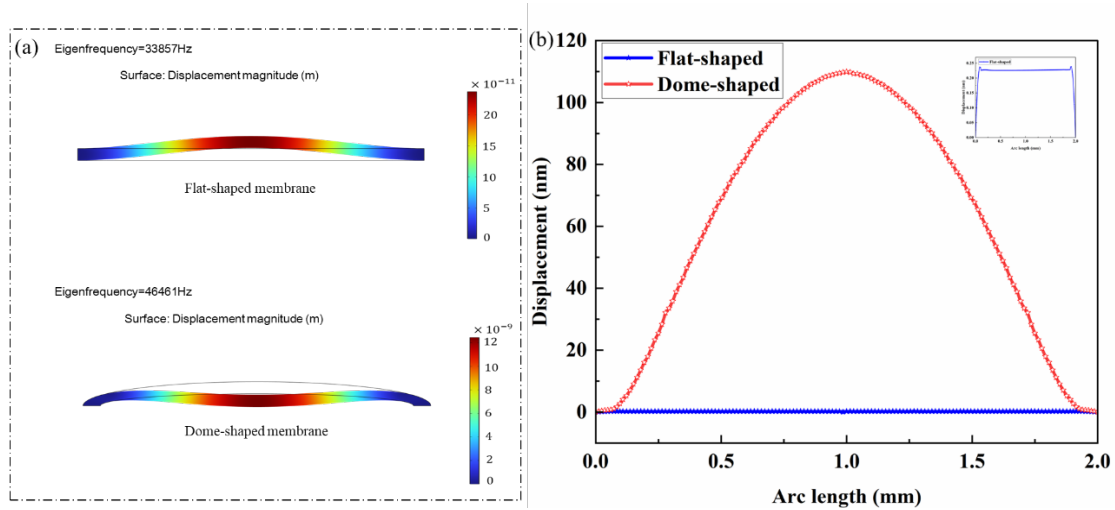

Fig.4 Simulated results of flat-shaped and dome-shaped PVDF films. (a) Deformations of flat and dome-shaped geometry. (b) Simulated displacement of the flat and dome-shaped geometry.

Furthermore, geometric parameters such as thickness, central height, and radius were systematically studied to evaluate their impact on acoustic performance metrics, including displacement amplitude and resonant frequency, the corresponding results are presented in Fig.5. Fig.5 (a) shows the relationship between displacement/resonant frequency and thickness for two dome radii (R = 0.5 mm and R = 1 mm). It is obvious that the displacement increases at the beginning then decreases significantly for both radii as the thickness increases. This is because thicker domes exhibit higher stiffness, reducing their ability to deform under the applied electric field. Conversely, the resonant frequency increases with the increase of thickness as a result of enhanced structural rigidity. The trade-off suggests that thinner films are favorable for applications requiring higher displacement, while thicker films are more suitable for higher-frequency operations. Fig.5 (b) examines the impact of central height on displacement and resonant frequency for two radii (R = 0.5 mm and R = 1 mm) but the same thickness. As the central height increases, the displacement initially rises, reaching a peak, and then gradually decreases. This behavior reflects the fact that an optimal central height enhances the strain distribution and displacement without overly increasing the dome's structural stiffness. The resonant frequency steadily increases with increasing central height, but it tends to decrease after it reaches the peak as the taller dome reduces overall stiffness. It's clear from Fig.5 (a) and (b) that the dome with smaller radii results in higher resonant frequencies, whereas the displacement exhibits the opposite trend, with larger radii producing greater displacement. Fig.5 (c) explores the effects of varying the dome radius on displacement and resonant frequency for two different thicknesses (T = 25 μm and T = 55 μm). As the radius increases, the displacement amplitude shows steadily increasing tendency. Larger radii improves the strain distribution, allowing for greater displacement, but excessively large radii may reduce the structural stability. The resonant frequency decreases consistently with increasing radius, as larger domes exhibit lower stiffness. This suggests that a moderate radius is ideal for balancing displacement and resonant frequency.

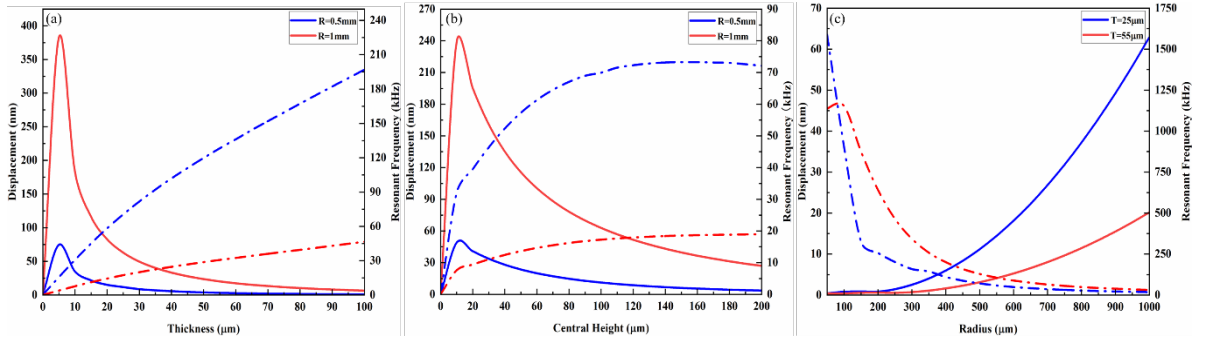

Fig.5 Vibration of ultrathin piezoelectric domes. (a) Surface displacement and resonance frequency vs film thickness. (b) Surface displacement and resonance frequency vs dome central height. (c) Surface displacement and resonance frequency vs dome radius.

## VI. Measurement of ultrathin PVDF transducer with micro-domes

### A. Acoustic performance of the flexible transducer

The acoustic performance of the flexible device, operating as an underwater transducer, was experimentally assessed using the setup depicted in Fig. 6. Fig. 6 (a) illustrates the schematic diagram of the acoustic testing setup, while Fig. 6 (b) displays the actual experimental setup. In this experiment, the transducer is driven by a sinusoidal AC voltage (20 $V_{pp}$), which is supplied by a RIGOL DG1022 function generator. To minimize overall bending during excitation, the flexible transducer is mounted on a rigid acrylic substrate. An Ocean Sonics RB9-ETH hydrophone is used to capture the acoustic signals, with a maximum measurable frequency of up to 200 kHz. The hydrophone is positioned 100 cm away from the transducer. The depth of the water in the tank is 30 cm.

Upon excitation, the microscale domes on the flexible transducer experienced mechanical deformation due to their intrinsic piezoelectric properties. This deformation converted the electrical input from the function generator into mechanical vibrations, producing acoustic waves in the surrounding water. These acoustic waves were subsequently detected by the hydrophone and converted into electrical signals proportional to the incident sound pressure. The sound pressure (SP) can be calculated using the following expression [32]:

$$SP = \frac{V}{S} \qquad (6)$$

Where $V$ represents the measured voltages and $S$ denotes the sensitivity of the hydrophone. The resulting sound pressure is then converted to SPL in decibels (dB) by formula (4).

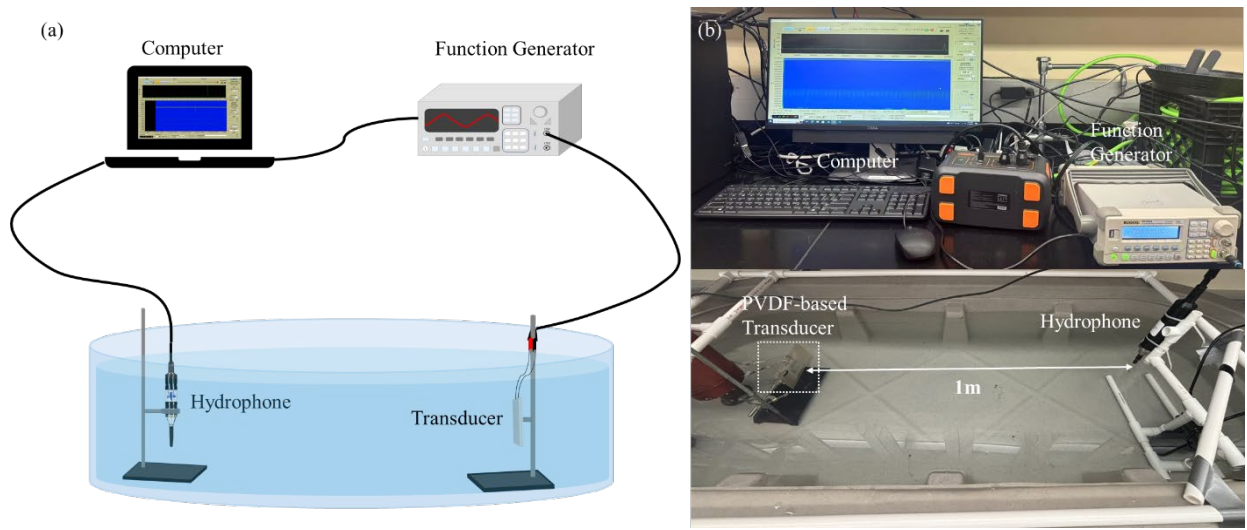

Fig.6 Experimental setup for evaluating the acoustic performance of the flexible transducer. (a) Schematic diagram of the acoustic testing setup. (b) Photograph of the actual experimental setup.

To investigate the impact of various design parameters on the acoustic performance of the flexible transducer, three sets of experiments were conducted. In all cases, the transducers were fabricated with a fixed overall size of 3 cm (length) × 3 cm (width), and each microscale dome featured a consistent radius of 1 mm. SPL was used as the primary metric to evaluate acoustic performance across different design conditions.

For the first experiment, flat-shaped and dome-shaped transducers with a thickness and central height of 100 μm were tested to investigate the geometric influence on the acoustic performance, the results are presented in Fig.7 (a). The SPL of the dome-shaped transducer remains above 92 dB throughout the tested frequency range of 10 kHz to 200 kHz, with peaks reaching 102 dB. In comparison, the flat transducer follows a similar general trend but shows lower amplitude across the frequency range, highlighting its limitations in efficiently generating acoustic output. This indicates that the dome-shaped geometry provides superior to the enhanced deformation and improved energy coupling with the surrounding water. Additionally, the dome-shaped transducer exhibits obvious improvements in SPL at higher frequencies (over 100 kHz), with more pronounced peaks in its frequency response, suggesting a significant resonance effect. In contrast, the flat-shaped transducer presents a relatively dampened response at higher frequencies, further emphasizing the advantages of the dome-shaped design.

In the second experiment, the dome-shaped transducers with three different thicknesses (25 μm, 55 μm, and 100 μm) but the same central height (100 μm) were tested to verify the influence of thickness on the acoustic performance of the flexible transducers. The results, depicted in Fig. 7 (b), show that the thinnest transducer (25 μm) consistently achieves the highest SPL across the frequency range of 10 kHz to 200 kHz. Its SPL peaks are the most pronounced, demonstrating a superior resonance effect and enhanced acoustic energy generation compared to the other two transducers. At lower frequencies (<100 kHz), the SPL differences among the transducers are less noticeable, although the transducer with a thickness of 25 μm holds a slight advantage. At higher frequencies (>100 kHz), the 25 μm transducer maintains its superior performance with higher SPL and sharper resonance peaks. In contrast, the flexible transducers with the thickness of 55 μm and 100 μm exhibit weak responses, particularly the 100 μm one. This is due to the increased mechanical stiffness of the thicker PVDF films, which limits their deformation and weakens the energy coupling with the surrounding water. As a result, the thicker transducers (55 μm and 100 μm) exhibit lower SPLs, while the thinner 25 μm transducer benefits from greater mechanical flexibility, allowing for more significant deformations and more efficient energy transmission, leading to superior acoustic performance.

For the third experiment, two flexible transducers with different central heights (H=100μm, H=200μm) but the same thickness (T=25 μm) were tested to study the dependence of the sound generation on the dome height. Fig.7 (c) shows that the transducer with a higher dome height (H = 200 μm) generally achieves higher SPL across the frequency range compared to the transducer with a central height of 100 μm. Specifically, 108 dB, 107 dB, 107 dB and 104 dB were produced at driving frequencies of $f$=20 kHz, 120 kHz, 150 kHz and 190 kHz, respectively. This suggests that the increased dome height contributes to greater displacement or mechanical strain in the transducer's structure, leading to a more efficient conversion of electrical energy to sound energy.

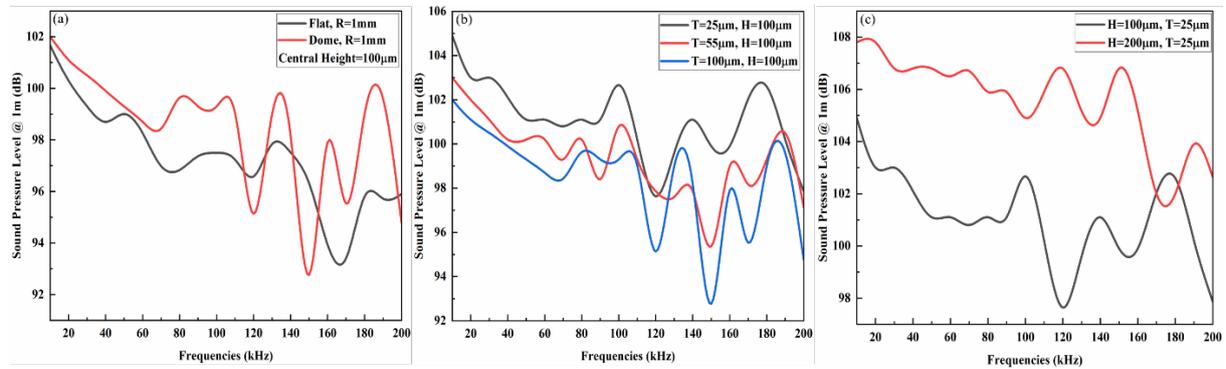

Fig.7 SPLs of ultrathin PVDF-based transducers with different design variables tested at 100 cm. (a) SPLs of flexible transducers based on two geometries with the same central height. (b) SPLs of flexible transducers with three different thickness but identical central height. (c) SPLs of flexible transducers with two different central heights but the same thickness.

The dependence of the first resonance frequency on the film thickness is studied by comparing the domes of the same size (R=1mm, H=100 μm) but different thicknesses (T=25 μm, 55 μm and 100 μm). Similarly, two films (R=1mm, T=25 μm) with different dome sizes (H=100 μm, 200 μm) are chosen to study the dependence of the resonance frequency on the dome central height. Fig.8 (a) and (b) shows reasonably good matching between the measured resonance frequencies of the domes obtained from the acoustic performance. Both the theoretical and experimental results present a similar trend that the first resonance frequency increases with the increase of the film thickness and the dome central height, which indicates the thin film with the large domes are more favorable for achieving high sensitivity in sound generation.

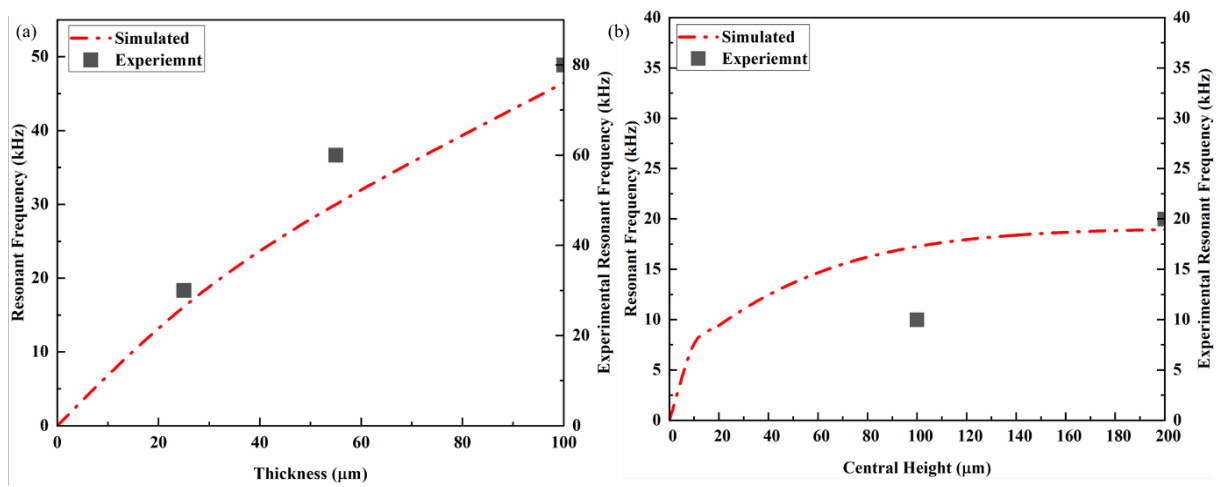

Fig.8 Frequency response of dome vibration. (a) Dependence of the first resonance frequency on the film thickness. (b) Dependence of the first resonance frequency on the dome central height.

## B. Underwater communication of the flexible transducer

To evaluate the performance of the microdome array-based PVDF transducer in underwater communication, the transducer was driven by a commercial modem (Subnero M25M series) at various voltage levels. The logo of the university of Alabama served as the transmitted signal, which was first digitized by MATLAB and adjusted to a bandpass range of 20 kHz to 30 kHz for transmission. The modem supported a data rate of up to 15 kbps and the corresponding modulation scheme was Frequency Hopping-Binary Frequency Shift Keying (FH-BFSK) [33], where the carrier frequency periodically shifts according to a predefined hopping pattern. The hydrophone (Ocean Sonics RB9-ETH), positioned 100 cm from the transducer, captured the acoustic signals and converted them into spectrograms using the Fast Fourier Transform (FFT) [34]. The corresponding schematic of the communication process and the corresponding experimental setup are demonstrated in Fig.9.

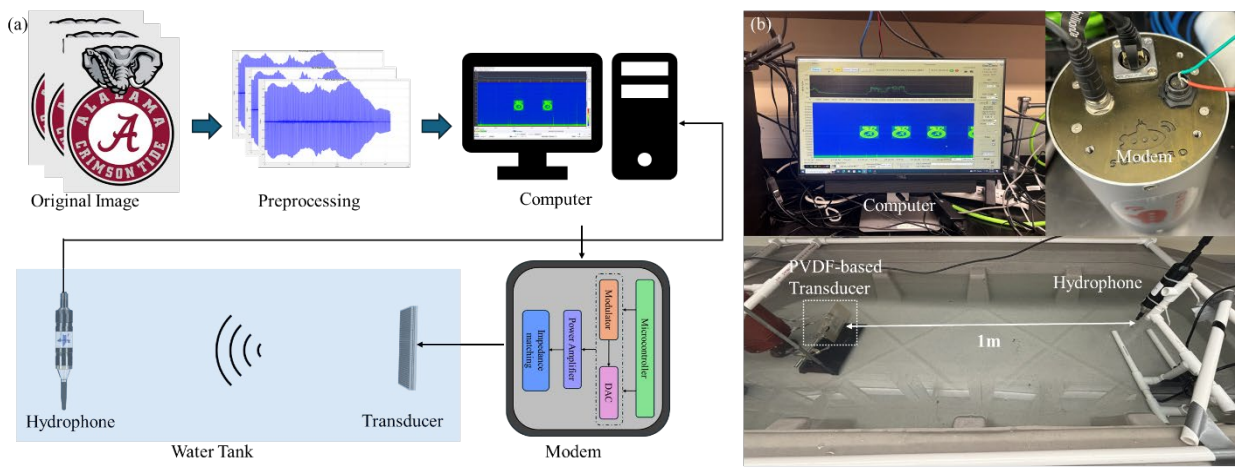

Fig.9 Underwater communication setup for the flexible transducer. (a) Schematic diagram of the experimental setup. (b) Photograph of the actual experimental setup.

Fig.10 presents underwater communication performance of the flexible PVDF-based transducer under varying transmission voltages. (a)–(d) shows the spectrograms of received signals at different voltage levels. And (e) illustrates signal-to-noise ratio (SNR) as a function of transmitter voltage. From Fig. 10 (a) to (d), a clear decrease in signal clarity and intensity is observed as the transmission power decreases. Fig.10 (e) indicates a strong positive correlation between transmission power and signal quality. As the transmitter voltage increases from –30 dB to 0 dB, a substantial improvement in SNR is observed, rising from approximately 5 dB to over 27 dB. These results demonstrate the ultrathin PVDF-based transducer's ability to enable visualizable underwater data transmission even at reduced voltage levels and further validates its effectiveness in underwater communication, which highlights its potential as a promising solution for low-power underwater communication systems.

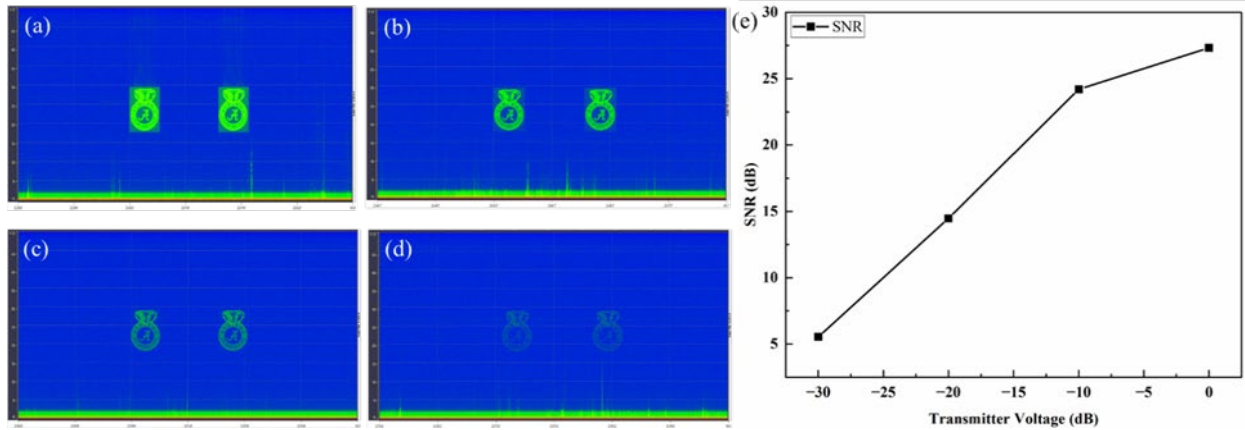

Fig.10 Underwater image transmission under different drive voltages at (a) 0 dB, (b) -10 dB, (c) -20 dB, (d) -30 dB. (e) SNR results of the collected acoustic signals.

## VII. Conclusion

In this study, we developed a flexible thin-film underwater transducer based on a mesoporous PVDF film embedded with microdome structures. The employment of ZnO NPs has been demonstrated as a successful method for the fabrication of sponge-like porous PVDF films. The experimental characterization of films matches well with the theoretical expectations that the enlarged porosity in PVDF film can enhance β-phase, improve mechanical compliance, and increase piezoelectric performance. The microstructures on the surface of the film are capable of vibrating independently of the mounting surface condition, enabling the transducer to operate effectively across a wide range of application scenarios. Additionally, this design can significantly enhance acoustic performance at high frequencies compared to the flat-shaped design. Furthermore, sound generation of the transducer can be boosted by increasing the dome size and reducing the film thickness. Such tunabilty of the transducer performance has been further verified through series of experiments. The information transmission experiment we conducted validates the ability of the transducer in reliably transmitting signals at low driving voltages. Additionally, it facilitates the visualization of underwater data communication over short distances, which means the proposed ultrathin PVDF-based transducer offers a promising solution for compact, low-power, and flexible underwater acoustic communication systems. In the future, we will focus our work on integrating the transducer into compact AUVs and incorporating machine learning techniques for advanced processing of the acquired signals.

## Acknowledgements

The authors would like to thank Fairoz Abida for her helpful guidance in PVDF film fabrication. The project was supported by NSF (1917462) and NOAA CIROH.## Reference

[1] N. Chaabane, S. Mahfoudhi and K. Belkadhi, "Interpolation-Based IoT Sensors Selection," in IEEE Sensors Journal, vol. 24, no. 21, pp. 36143-36147, 1 Nov.1, 2024.